\documentclass[aps,prx,reprint,superscriptaddress,noeprint,showkeys]{revtex4-2}

\usepackage[hidelinks]{hyperref}
\usepackage{xcolor}
\hypersetup{
    colorlinks,
    linkcolor={blue},
    citecolor={blue},
    urlcolor={blue},
}
\usepackage{graphicx}
\usepackage{amsmath,braket}

\newcommand{\CrBr}{$\mathrm{CrBr_3}$}
\newcommand{\Tc}{$T_c$}
\newcommand{\Xac}{$\chi_{AC}$}

\begin{document}


\title{AC susceptometry of 2D van der Waals magnets enabled by the coherent control of quantum sensors}

\author{Xin-Yue Zhang}
\author{Yu-Xuan Wang}
\author{Thomas A. Tartaglia}
\author{Thomas Ding}
\author{Mason J. Gray}
\author{\\Kenneth S. Burch}
\author{Fazel Tafti}
\author{Brian B. Zhou}
\email{Correspondence to: brian.zhou@bc.edu}
\affiliation{Department of Physics, Boston College, Chestnut Hill, MA, 02467, USA}

\date{\today}

\begin{abstract}
Precision magnetometry is fundamental to the development of novel magnetic materials and devices. Recently, the nitrogen-vacancy (NV) center in diamond has emerged as a promising probe for static magnetism in 2D van der Waals materials, capable of quantitative imaging with nanoscale spatial resolution. However, the dynamic character of magnetism, crucial for understanding the magnetic phase transition and achieving technological applications, has rarely been experimentally accessible in single 2D crystals. Here, we coherently control the NV center's spin precession to achieve ultra-sensitive, quantitative ac susceptometry of a 2D ferromagnet. Combining dc hysteresis with ac susceptibility measurements varying temperature, field, and frequency, we illuminate the formation, mobility, and consolidation of magnetic domain walls in few-layer \CrBr. We show that domain wall mobility is enhanced in ultrathin \CrBr{}, with minimal decrease for excitation frequencies exceeding hundreds of kilohertz, and is influenced by the domain morphology and local pinning of the flake. Our technique extends NV magnetometry to the multi-functional ac and dc magnetic characterization of wide-ranging spintronic materials at the nanoscale.

\end{abstract}
\maketitle

\section{Introduction}
Two-dimensional van der Waals (vdW) magnetic materials have broached transformative concepts for integrating and controlling spintronic devices, but also presented unique challenges to their magnetic characterization \cite{Burch2018a,Gong2019,Mak2019}. Notably, the magnetic moment of a micron-sized exfoliated monolayer (${\sim}10^{-11}$ emu) lies below the sensitivity (${\sim}10^{-8}$ emu) of commercial magnetometers based on the superconducting quantum interference device (SQUID) \cite{Buchner2018}, the benchmark for analyzing bulk magnetic materials. Key insights in vdW magnetism are most frequently provided by magneto-optical \cite{Huang2017,Gong2017,Jin2019,Zhang2020} and photoluminescence (PL) imaging \cite{Seyler2018,Zhang2019}, which feature high sensitivity, or by electrical measurements, which directly incorporate layers inside functional devices \cite{Ghazaryan2018,Klein2018,Wang2019b}. However, these approaches depend sensitively on material- and device-specific coupling of the sample's magnetization to the measured optical or transport properties, which complicates quantitative interpretation and restricts generalizability \cite{Wu2019a,Paudel2019,Molina-Sanchez2020}. A direct probe of the magnetic field associated with both static and dynamic magnetizations in 2D crystals would accelerate materials discovery and optimization.

Detecting stray fields from single 2D magnetic flakes is challenging due to the requirement of miniaturized sensors that must be deployed in nanoscale proximity to these microscopic samples. Pioneering efforts have utilized mesoscopic graphene-based Hall bars \cite{Kim2019} and atomic-scale spin defects in diamond \cite{Thiel2019,Broadway2020a,Sun2020,Fabre2020} to measure static stray fields and quantify the absolute magnetization of mono- and few-layer ferromagnets. However, the dynamic signatures of 2D magnets \cite{Jin2019,Zhang2020,Ghazaryan2018,Cenker2021}, which predominate near magnetic criticality, have yet to be investigated directly through magnetometry. In particular, ac susceptibility, which probes the response of the sample to an oscillating magnetic field, is a powerful technique for understanding magnetic phase transitions, relaxation times, and domain dynamics in diverse materials, including antiferromagnets, spin glasses, and ensembles of single-molecule magnets \cite{Baanda2013,Topping2019}. At the Curie temperature (\Tc) of the paramagnetic to ferromagnetic phase transition, the ac susceptibility $\chi_{AC}$ diverges with a critical exponent $\gamma$ characteristic of the universality class of the underlying interactions \cite{Gibertini2019}. Below \Tc, $\chi_{AC}$ senses magnetization rotation, domain wall motion, superparamagnetism, and the interactions of these processes with defects, strain, and external dc field \cite{Chikazumi1997}.

In this work, we leverage the coherent manipulation of quantum sensors to probe both the dc and ac magnetic properties of a 2D vdW magnet. We exfoliate ultrathin flakes of the ferromagnetic insulator \CrBr{} onto a diamond magnetometer chip containing a near-surface ensemble of nitrogen-vacancy (NV) centers. This versatile platform directly enables sensitive measurement of both the 2D ferromagnet's dc magnetization through static shifts of the NV center's spin energies \cite{Thiel2019,Broadway2020a,Sun2020,Fabre2020} and its ac susceptibility through dynamically coupling the NV center's spin precession to ac fields using frequency-selective quantum control sequences \cite{Lovchinsky2016,Zhou2019}. Our dc measurements of magnetic hysteresis reveal that the details of domain nucleation and pinning can differ significantly in ultrathin, exfoliated \CrBr{} flakes of similar thickness (${\sim}$10 layers), pointing to large variability in the impact of the local microstructure. Moreover, ac susceptibility measurements on few-layer \CrBr{} indicate a critical exponent $\gamma = 1.1 \pm 0.3$ for the ferromagnetic phase transition and display distinct features that illuminate the formation, mobility, and consolidation of single domain walls below $T_c$, informing on the potential for domain-based memory and logic devices \cite{Parkin2015,Velez2019,Grollier2020}. The remarkable ac field resolution of ${\sim}$40 nT achieved here represents the most sensitive magnetometry performed to date on exfoliated 2D magnets \cite{Kim2019,Thiel2019,Broadway2020a,Sun2020,Fabre2020} and introduces a generic platform for understanding sub-gigahertz dynamical phenomena in 2D magnetism.

\section{Magnetic Properties Measurement System for 2D vdW Magnets}
\begin{figure}[ht]
\includegraphics[scale=1]{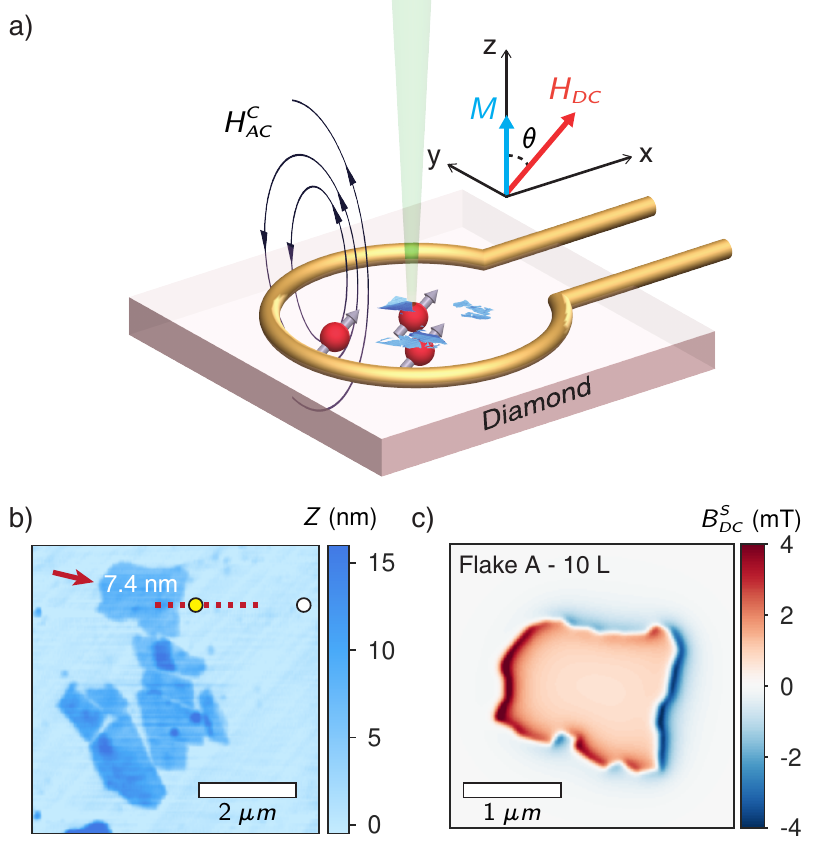}
\caption{\label{fig:1}Experimental overview. (a) \CrBr{} flakes (blue) are exfoliated onto a diamond chip containing a shallow ensemble of NV centers (red). A coil delivers both the ac field $H^C_{AC}$ for exciting the magnetic dynamics of the ferromagnet and the gigahertz-frequency pulses to manipulate the NV center spin. The static bias field $H_{DC}$ is aligned along the NV center axis, while the easy-axis for the magnetization $M$ is along the out-of-plane, $z$-direction. (b) AFM image of Flake A (arrow), 10 layers thick. Measurements are primarily taken outside the flake's right edge (yellow dot) and are referenced to the background signal far away from the flake (e.g., white dot). (c) Simulated stray field map for Flake A, assuming an NV center depth of 60 nm. $B^S_{DC}$ is the component of the flake's stray field parallel to the NV center axis.}
\end{figure}

Figure \ref{fig:1}a displays our experimental platform based on the optical readout of a layer of near-surface NV center spins ($\sim$60 nm depth). We exfoliate \CrBr{} flakes \cite{Abramchuk2018,Tartaglia2020} without encapsulation onto a diamond substrate inside an argon-filled glovebox and then transfer the substrate into our cryostat with minimized exposure to ambient (see Supplemental Material Sec. I). An insulated wire coil adjacent to the diamond delivers both the microwave pulses ($\sim$GHz) for NV center spin manipulation and the radio frequency ($\sim$100 kHz) excitation field $H_{AC}^C$ for probing the ac response of the ferromagnet. In this paper, we use $H$ to denote the magnitude of a magnetic field (e.g., $H^C_{AC}$), and $B$ to denote its projection onto the NV center axis (e.g., $B^C_{AC}$). The external magnetic field $H_{DC}$ is carefully aligned along one of the four NV center crystallographic orientations, forming an angle $\theta = 54.7^\circ$ to the surface normal. Away from \Tc, the applied, in-plane component of $H_{DC}$ in our experiments ($<$0.05 T) can be neglected as \CrBr{} possesses a large uniaxial anisotropy that pins the magnetization $M$ to the out-of-plane direction ($z$-axis) \cite{Kim2019a}.

We identify \CrBr{} flakes as thin as 6 layers for low-temperature measurement by their optical contrast on diamond. The thicknesses of the investigated flakes are subsequently verified by atomic force microscopy (AFM), as exemplified in Fig. \ref{fig:1}b, showing an image of a 7.4 nm-thick \CrBr{} flake (Flake A). In Fig. \ref{fig:1}c, we perform simulations of Flake A's stray field at 60 nm beneath the diamond surface. We assume a saturated magnetic moment density of 148 $\mu_b/\mathrm{nm}^2$ (3.0 $\mu_b$ per Cr ion) for 10 layers with ferromagnetic interlayer coupling and plot $B_{DC}^S$, the projection of the sample's stray field onto the oblique NV center axis \cite{Thiel2019}. In this work, we utilize the $\ket{m_S=0}\rightarrow\ket{m_S=-1}$ transition in the NV center ground state, which disperses with a slope $-\gamma_e/2\pi = -28$ MHz/mT \cite{Zhou2019}. Hence, outside the right edge of the flake, where many of our measurements are performed, we expect a strong positive (negative) Zeeman shift of the NV center's transition frequency if the flake is magnetized parallel (anti-parallel) to the component of $H_{DC}$ along the $z$-axis.

\section{DC field sensing of magnetic hysteresis}
\begin{figure*}[ht]
\includegraphics[scale=1]{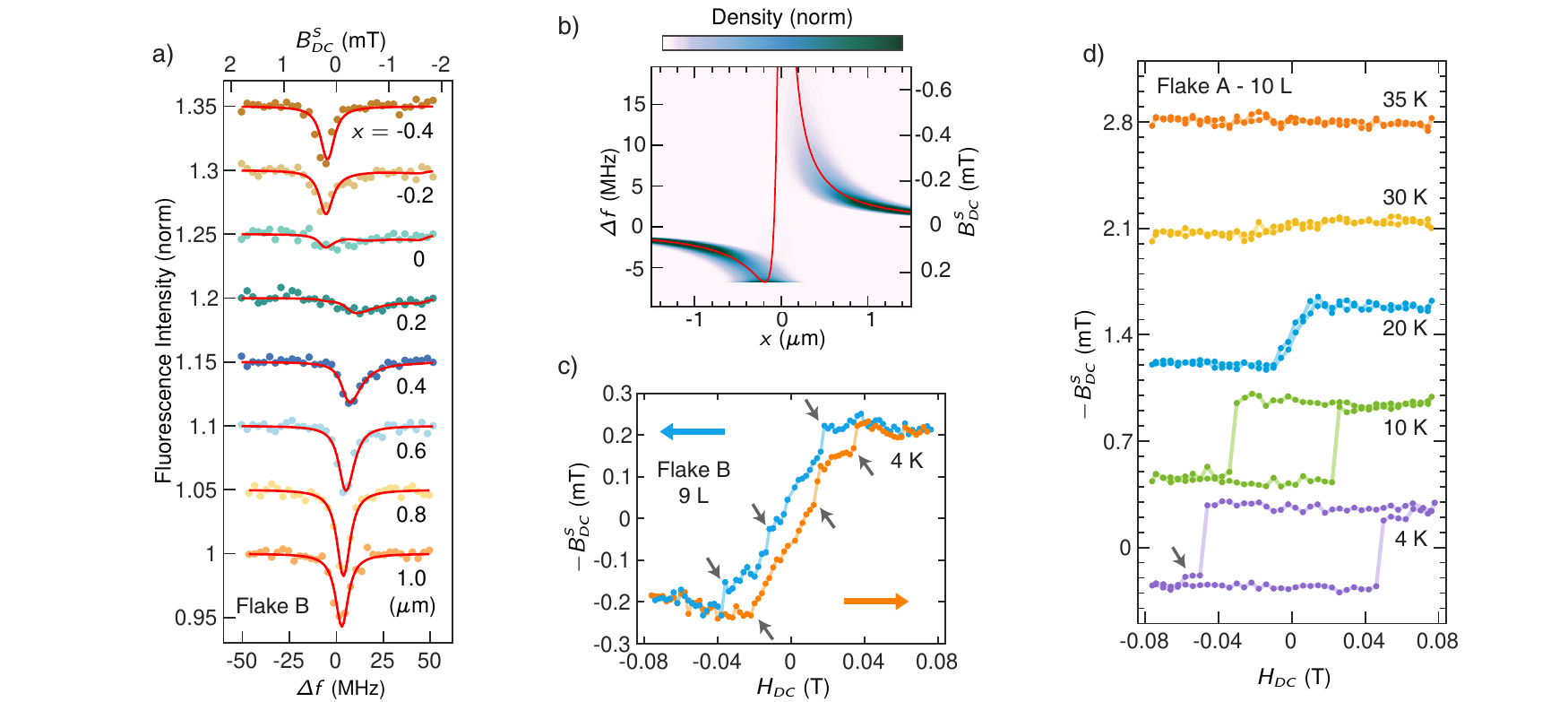}
\caption{\label{fig:2} DC magnetometry of ultrathin \CrBr{}. (a) Linecut of the ODMR spectra across the right edge of Flake B (9 layer) at $H_{DC}$ = 76 mT and $T$ = 4 K. Positive (negative) $x$ denotes the exterior (interior) of the flake. $\Delta f$ is the shift in frequency from the resonance of NV centers far from the flake ($x \approx 4\,\mu$m). The red lines are fits using a model that considers magnetic field gradients over the optical spot for an ensemble NV sample. (b) Illustration of the distribution of stray fields $B^S_{DC}$ over the optical spot sensed by the NV ensemble. The stray field distribution (density plot) at position $x$ corresponds to the stray field from an infinite magnetized edge (red line) weighted by a Gaussian distribution centered around $x$ for the optical intensity. (c) Magnetic hysteresis curve for Flake B. The magnetization $M$, which is proportional to $-B^S_{DC}$, slowly reverses due to domain wall pinning. Discrete Barkhausen jumps are marked by the arrows. (d) Temperature-dependent hysteresis curves for Flake A (10 layer). Flake A displays rectangular hysteresis, indicating minimal pinning of domain walls once nucleated.}
\end{figure*}

In Fig. \ref{fig:2}a, we present the optically-detected magnetic resonance (ODMR) spectra at 4 K for a linecut across the right edge of another \CrBr{} flake (Flake B, 6.7 nm/9 layers) with a long, straight edge (see Supplemental Material Sec. II for AFM image). Approaching the right edge of the flake from the outside ($x > 0$), the NV magnetic resonance shifts to higher frequencies, before switching to lower frequencies in the interior of the flake ($x < 0$), corroborating a ferromagnetic moment aligned with the out-of-plane component of $H_{DC}$. In comparison to single NV measurements \cite{Thiel2019}, we observe broadening of the ensemble NV center linewidth due to spatial averaging over strong magnetic field gradients near the flake's edge. The evolution of the lineshape, however, can be accurately modeled (solid lines in Fig. \ref{fig:2}a) by summing intrinsic Lorentizian lineshapes with a distribution of center frequencies (see Supplemental Material Sec. III). For each center location $x$ of the optical spot, we determine the distribution of stray fields $B_{DC}^S$ within the spot (Fig. \ref{fig:2}b) by assuming a Gaussian beam shape (standard deviation $\sigma_B =$ 170 nm) and a stray field profile corresponding to the edge of a semi-infinite magnetic sheet \cite{Hingant2015}. Fitting the lineshapes simultaneously using this model, we extract the magnetization of Flake B to be $86 \pm 5\;\mu_b/\mathrm{nm}^2$. This value is lower than expected for 9-layer \CrBr{} at saturation (133 $\mu_b/\mathrm{nm}^2$), and may be due to degradation of the surface layers during sample transfer or laser irradiation, as also observed in other NV center experiments \cite{Broadway2020a,Fabre2020}.

By parking the optical spot outside the flake, we demonstrate sensitive characterization of magnetic hysteresis in ultrathin \CrBr{} (see Supplemental Material for data on additional flakes) \cite{Zhang2020}. Figure \ref{fig:2}c displays the stray field $B_{DC}^S$ due to Flake B as the applied field $H_{DC}$ is swept, sensing the magnetization of a local region of the flake near the illuminated NV location. Lowering $H_{DC}$ from $76$ mT, we observe that the magnetization begins to reverse at $20$ mT through the nucleation of domains. The magnetization decreases linearly with field and crosses zero magnetization slowly, revealing that a highly fragmented domain structure is stable and that domain wall pinning impedes the expansion of the reversed domains. Discrete Barkhausen jumps (gray arrows) are observed, corresponding to the hopping of individual domain walls between different pinning sites \cite{Kim2019,Sun2020}. Interestingly, in contrast to the ``soft'' magnetic behavior of Flake B (9 layer), the hysteresis curve for a similarly thick flake, Flake A (10 layer), is rectangular, displaying full remenance and a large coercive field (Fig. \ref{fig:2}d). Only a single magnetization plateau is observed between the two saturated states, indicating a propensity to form large domains and few pinning centers in the vicinity of the NV center probe. These observations indicate that the metastable domain structures and magnetization reversal process in intermediate thickness \CrBr{} flakes (9-10 layer) are extremely sensitive to details of the local microstructure. As the temperature increases, the hysteresis loop for Flake A narrows, and the magnetization disappears above 30 K.

\begin{figure*}[ht]
\includegraphics[scale=1]{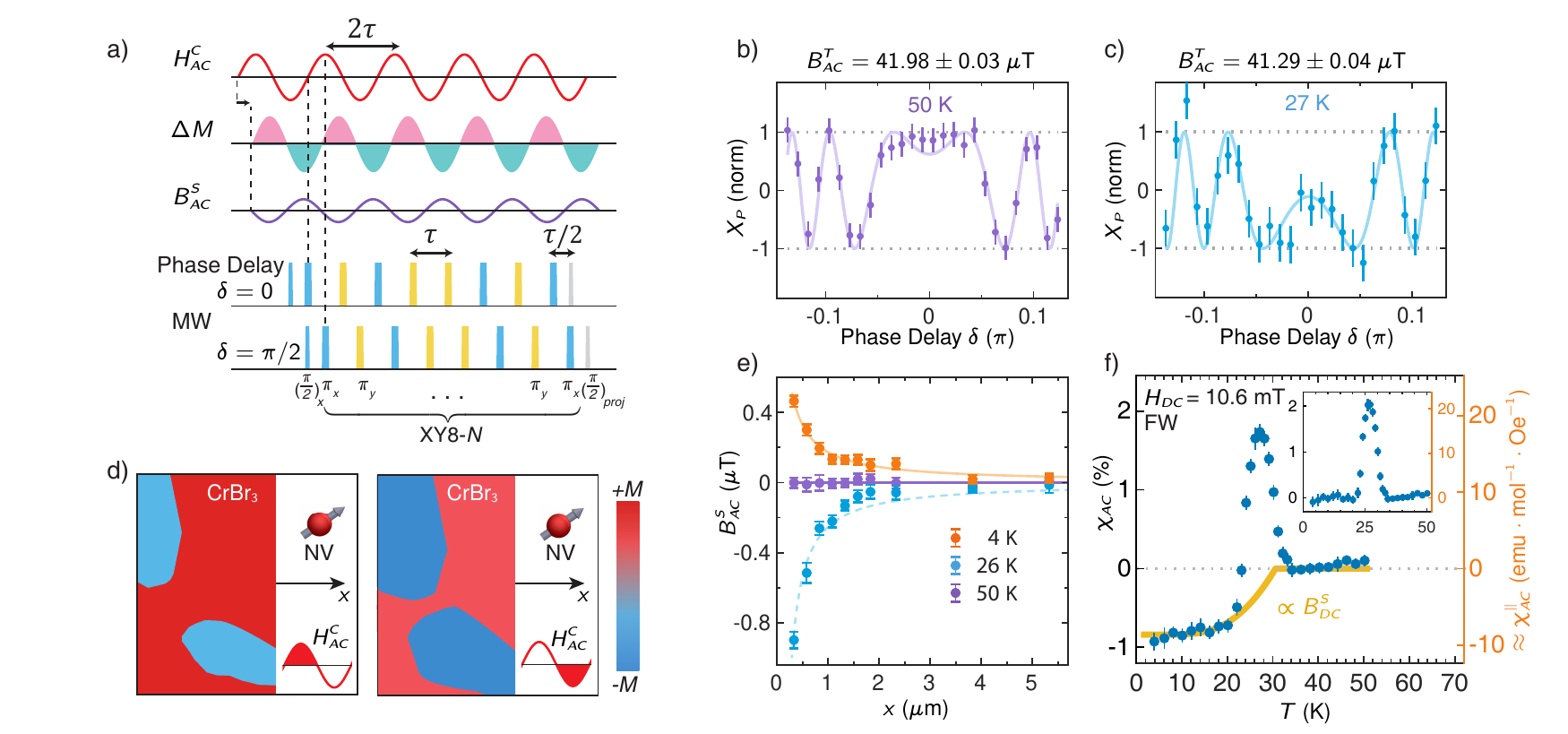}
\caption{\label{fig:3}AC susceptometry technique. (a) Experimental sequence: an oscillating coil field $H^C_{AC}$ excites an ac magnetization change $\Delta M$ (with a possible phase shift) in the \CrBr{} flake. An XY8-$N$ dynamical decoupling sequence is applied to the NV center to lock-in to the total ac field $B^T_{AC} = B^C_{AC}+B^S_{AC}$ due to both the coil and the sample. The phase delay $\delta$ between $H^C_{AC}$ and the microwave (MW) pulses can be tuned to isolate either the in-phase ($\delta = 0$) or out-of-phase ($\delta = \pi/2$) magnetization response. (b) $B^T_{AC}$ at 50 K for Flake A determined from the fringe pattern around $\delta = 0$ in the $x$-projection, $X_P$, of the NV superposition state after precession due to the ac fields. The data were acquired at $H_{DC}$ = 10.6 mT using an XY8-$N2$ sequence with $f_{AC}$ = 238 kHz. (c) $B^T_{AC}$ at 27 K. (d) Schematic showing the microscopic mechanisms for $\Delta M$. $H^C_{AC}$ causes domain wall displacements or magnetization rotation in the sample. (e) Linecut of the sample's ac stray field $B^S_{AC}$ $(= B^T_{AC}-B^C_{AC})$ away from the flake's right edge. A linear background for the coil field $B^C_{AC}$ is determined from the 50 K data and is subtracted from all three data sets. The positive $B^S_{AC}$ at 4 K is an artifact induced by the flake's dc stray field $B^S_{DC}$. (f) Temperature dependence of the uncorrected ac susceptibility \Xac{} during field-warming (FW) from a saturated state. The yellow curve is the baseline for the susceptibility and is proportional to $B^S_{DC}$. Inset: Baseline-corrected \Xac{} due to the physical signal.}
\end{figure*}

\section{Quantum-Enabled AC susceptometry technique}
We now introduce the main innovation of our work: ac susceptometry via spin precession-based ac field sensing. Our protocol utilizes dynamical decoupling (DD) sequences (XY8-\emph{N}) that consist of repeated spin inversions ($\pi$-pulses) on an NV superposition state undergoing Larmor precession \cite{Degen2016}. Similar to lock-in detection, the DD sequence allows only ac fields synchronized to the $\pi$-pulse spacing to affect the NV precession, while blocking the influence of wideband magnetic noise. This approach extends the coherence time ($T_2$) of the NV superposition state and significantly enhances the magnetic field sensitivity compared to dc (ODMR) measurements, enabling detection of subtle phenomena in proximal systems, such as nuclear magnetic resonance \cite{Lovchinsky2016,Aslam2017,Glenn2018}, electron paramagnetic resonance \cite{Shi2015}, and photocurrent transport \cite{Zhou2019}. As outlined in Fig. \ref{fig:3}a, an oscillating coil field with amplitude $B^C_{AC}$ along the NV center axis and frequency $f_{AC}$ ($\sim$200 kHz) induces an ac magnetization response $\Delta M$ in the sample, which produces its own ac stray field $B^S_{AC}$. After an applied DD sequence with pulse spacing $\tau = 1/(2f_{AC})$, the NV center precesses by a cumulative angle $\phi_{NV}$ due to the total resonant ac field parallel to the NV axis, $B^T_{AC} = B^C_{AC} + B^S_{AC}$, containing contributions from both the coil and the sample.

The ac susceptibility $\chi_{AC} = dM/dH$ is in general complex-valued, with a real (imaginary) part from the in-phase (out-of-phase) sample response. Our DD sequence is phase-sensitive by controlling the relative timing, denoted as the phase delay $\delta$, between the applied coil field and the train of $\pi$-pulses applied to the NV center. For $\delta = 0$, the NV precession detects $B^C_{AC} + \operatorname{Re}(B^S_{AC})$, while for $\delta = \pi/2$, it detects $\operatorname{Im}(B^S_{AC})$ (Fig. \ref{fig:3}a). Our measurement scans $\delta$ about $\delta=0$ to map a fringe pattern in the $x$-projection, $X_P = \cos(\phi_{NV})$, of the final NV superposition state on the Bloch sphere. Fitting this fringe pattern allows us to robustly determine $B^T_{AC}$ and resolve $2\pi$ ambiguities in the NV's acquired phase (see Supplemental Material Sec. IV). For the small amplitude excitation fields applied, we do not detect any out-of-phase response in \CrBr{} flakes as thin as 10 layers within our experimental sensitivity. Hence, $\chi_{AC}$ in our paper denotes solely the real part of the susceptibility, reflecting reversible magnetization changes. The absence of an imaginary component, due to dissipative or irreversible processes, is consistent with ac magneto-optic Kerr effect (ac-MOKE) measurements on multi-layer \CrBr{} flakes at much lower $f_{AC}$ = 36 Hz, although an imaginary component was observed for monolayer \CrBr{} \cite{Jin2019}.

We first calibrate our measurement by detecting $B^T_{AC}$ near the right edge of Flake A at $T$ = 50 K and $H_{DC}$ = 10.6 mT (Fig. \ref{fig:3}b). Far above \Tc{} of \CrBr{}, $B^T_{AC}$ is contributed only by the coil, which we determine to produce a peak amplitude of $41.98 \pm 0.03\,\mu$T along the NV center axis. Over the 33.6 $\mu$s duration of the XY8-$\it{N2}$ sequence used here, the coil field induces a large maximum precession angle $\phi_{NV} \approx 50\pi$ for $\delta = 0$ and leads to rapid oscillations in $X_P$ for small changes in $\delta$.  Tracking the same spot relative to the flake's edge, we find that $B^T_{AC}$ is reduced to $41.29 \pm 0.04\,\mu$T at 27 K (Fig. \ref{fig:3}c). We verify that the coil field $B^C_{AC}$ does not change with temperature by measuring NV centers far from any \CrBr{} flakes (Supplemental Material). Hence, we conclude that Flake A produces an ac stray field $B^S_{AC} \approx -0.69\;\mu$T, which is remarkably $>$400 times smaller than its dc stray field at saturation (Fig. \ref{fig:2}d).

As illustrated Fig. \ref{fig:3}d, the susceptibility $\chi_{AC}$ is contributed microscopically by both magnetization rotation (MR) and domain wall displacement (DWD). For MR, the coil field $H^C_{AC}$ exerts a torque that aligns sample spins with variable orientations along the direction of $H^C_{AC}$, here predominantly along the $z$-axis, also the easy-axis for \CrBr. Alternatively for DWD, domains aligned with $H^C_{AC}$ grow in area at the expense of anti-aligned domains. Hence, for both MR and DWD, $\Delta M$ of the flake is positively correlated to $H^C_{AC}$. This implies that the sample field $B^S_{AC}$ should indeed have opposite sign to the coil field $B^C_{AC}$ outside the flake's right edge (cf. Fig. \ref{fig:1}c).

Mapping $B^S_{AC}$ versus $x$ in Fig. \ref{fig:3}e, we corroborate that Flake A's ac stray field is negative at 26 K and decays away from the flake's edge; however, a surprising positive signal is observed at 4 K. The linecuts in Fig. \ref{fig:3}e are obtained during field-warming (FW) at $H_{DC}$ = 10.6 mT after fully polarizing the flake with $H_{DC}$ = 76 mT. In this case, $\chi_{AC}$ should be zero starting in the saturated, single domain state at 4 K. The observed positive signal is a measurement artifact that is induced by the flake's dc stray field $H^S_{DC}$, which introduces a variable coupling to the component of the coil field $H^C_{AC}$ perpendicular to the NV axis. As the flake is approached, the total dc field $H_{DC}+H^S_{DC}$ becomes misaligned from the NV center axis, and the NV spin precession is affected not only by $B^C_{AC}$ (parallel to the NV axis), but also to a lesser degree by the component of $H^C_{AC}$ perpendicular to the NV axis. This ${\sim}1\%$ systematic deviation is resolved by the high precision of our measurements. We replicate this effect by deliberately translating our permanent magnet, which supplies $H_{DC}$, to show that the accumulated phase $\phi_{NV}$ scales linearly in small dc fields perpendicular to the NV axis, in agreement with theoretical analysis (see Supplemental Material Sec. V).

Accordingly, the spatial dependence of the artifact $B^S_{AC}$ at 4 K fits accurately to the decay (orange line) of the dc stray field perpendicular to the NV center axis due to the edge of a homogeneous, out-of-plane magnetized sheet \cite{Hingant2015}. On the other hand, the spatial dependence $B^S_{AC}$ at 26 K, containing physical signals from ac magnetization changes inside the flake, displays a faster decay that is better approximated by the stray field parallel to the NV axis from the same magnetic edge. This fit (dashed blue line) may be imperfect, for example if the spatial distribution of domain walls is not homogeneous.

To demonstrate how to extract the physical ac susceptibility signal, we present the full temperature dependence of $B^S_{AC}$ in Fig. \ref{fig:3}f. The data here are measured at fixed location ($x \approx 0.33\,\mu$m) during field-warming Flake A from the saturated state. We define a subjective metric, $\chi_{AC} \equiv -B^S_{AC}/B^C_{AC}$, where $B^S_{AC}$ depends on the measurement location and the temperature-independent coil field $B^C_{AC}$ is determined by measurements above \Tc{} (Fig. \ref{fig:3}b). Since the spurious contribution to $B^S_{AC}$ is proportional to the dc field perpendicular to the NV axis, we should subtract a baseline proportional to the temperature-dependent dc magnetization of the flake. For each temperature during the sweep, we simultaneously sample the dc magnetization by measuring $B^S_{DC}$ via ODMR (see Supplemental Material for data). The baseline for \Xac{} (yellow curve in Fig. \ref{fig:3}f) is thus determined by scaling $B^S_{DC}$ by a multiplicative factor to match the raw $\chi_{AC}$ data for $T <$ 10 K, where the physical susceptibility should be zero since Flake A remains saturated based on the dc magnetization.

The baseline-subtracted $\chi_{AC}$, displayed in the inset of Fig. \ref{fig:3}f, reveals the onset of susceptibility via domain nucleation upon warming to $T = 20 $ K. $\chi_{AC}$ peaks at 27 K, below $T_c = 30.5 \pm 0.5$ K of Flake A (determined by field-cooling (FC) initial susceptibility measurements presented in the next section). A peak in $\chi_{AC}$ below $T_c$ is commonly observed in ferromagnets and is known generically as a Hopkinson peak, which can stem from different physical origins \cite{Baanda2013,Salas1990}. The Hopkinson peak for the FW trace here originates from an increase in domain wall density and mobility as the temperature increases, which is ultimately counterbalanced by the loss of saturation magnetization $M_S$ towards \Tc.

To convert our data to quantitative units, we estimate the molar susceptiblity, $\chi^\parallel_{AC}$, by accounting for the distance between the NV probe location and the flake's edge and for the angle of flake's edge with respect to the NV axis.  By necessity, we disregard the spatial details of the source of susceptibility and consider the equivalent magnetization change, $\Delta M$, of a semi-infinite, homogeneous flake that would produce the same $B^S_{AC}$ at the NV location (see Supplemental Material Sec. VI). Dividing $\Delta M$ by the amplitude of $H^C_{AC}$ along the $c$-axis of \CrBr{}, we obtain $\chi^\parallel_{AC}$ (Fig. \ref{fig:3}f, secondary y-axis), where we assume the component of $H^C_{AC}$ perpendicular to the easy axis makes a negligible contribution, which is valid for DWDs under strong magnetic anisotropy. During FW, the maximum $\chi^\parallel_{AC}$ for Flake A is 20 emu/(mol Oe) for ${\sim}40\;\mu$T excitation. For reference, a peak $\chi^\parallel_{AC} \approx 6$ emu/(mol Oe) was measured for bulk $\mathrm{CrI_3}$ using a SQUID magnetometer \cite{Liu2018a}. While SQUID measurements treat the sample as a point dipole, our $\chi^\parallel_{AC}$ should be taken as a local measurement that may vary along the length of the flake.

\section{Initial AC Susceptibility}
\begin{figure*}[ht]
\includegraphics[scale=1]{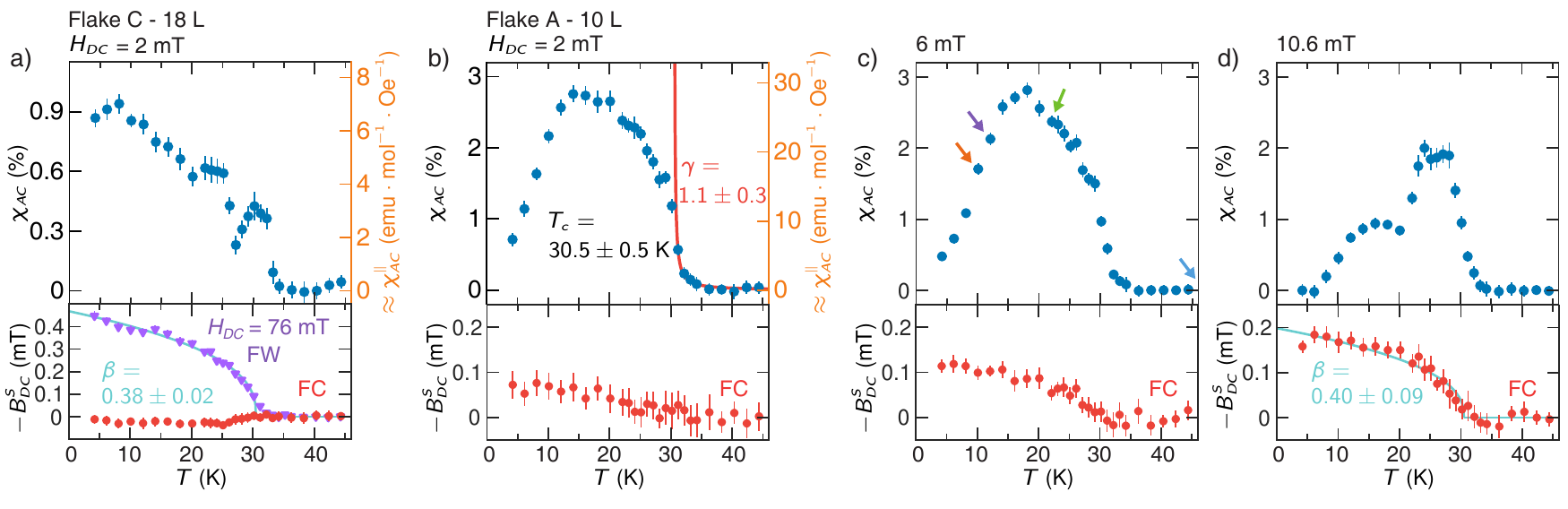}
\caption{\label{fig:4}Initial ac susceptibility for bulk-like and ultrathin \CrBr{}. (a) \Xac{} (blue) for bulk-like Flake C (18-layer) during field-cooling (FC) at low bias field $H_{DC}$ = 2 mT. The left y-axis, \Xac{}, is defined as $-B^S_{AC}/B^C_{AC}$, while the right y-axis, $\chi^\parallel_{AC}$, is a quantitative estimate for the molar susceptibility that accounts for the distance and angle of the flake's edge relative to the NV probe. The dc magnetization (red), proportional to $-B^S_{DC}$, remains zero during FC, indicating a finely fragmented domain structure. The magnetization during field-warming (FW, purple) at a polarizing field $H_{DC}$ = 76 mT follows a critical scaling $M(T) \propto (1 - T/T_c)^\beta$ with $\beta = 0.38 \pm 0.02$. (b-d) \Xac{} (blue) and $B^S_{DC}$ (red) for Flake A (10 layer) during FC at (b) $H_{DC}$ = 2 mT, (c) 6 mT, and (d) 10.6 mT. Flake A displays rectangular hysteresis and larger domains more common for ultrathin \CrBr{}. \Xac{} diverges at \Tc{} = 30.5 K with a critical exponent $\gamma = 1.1 \pm 0.3$ (red fit in (b)), while the dc magnetization scales with a critical exponent $\beta = 0.40 \pm 0.09$ (cyan fit in (d)). The arrows in (c) indicate temperatures for frequency-dependent measurements presented in Fig. \ref{fig:5}. All \Xac{} data here are taken at $f_{AC}$ = 238 kHz with XY8-$N2$ and are baseline-corrected using the flake's dc stray field $B^S_{DC}$.}
\end{figure*}

Having detailed our experimental technique and analysis, we now present baseline-corrected measurements of $\chi_{AC}$ for few-layer \CrBr{} during the initial magnetization process from the paramagnetic state. We begin with Flake C, a thicker 18-layer flake, that displays a hysteresis loop with minimal opening area at 4 K (see Supplemental Material), approaching the linear retraceability observed for bulk samples \cite{Abramchuk2018}. As shown in Fig. \ref{fig:4}a (red points), the dc stray field $B^S_{DC}$ for Flake C remains virtually vanishing throughout FC at small $H_{DC}$ = 2 mT, indicating the stabilization of an equally-balanced, fine-grained domain structure. Indeed, electron microscopy observes periodic stripe-like domains in bulk \CrBr{} with sub-micron widths \cite{Bostanjoglo1970a}, consistent with significantly reduced stray fields even when measured at our near-field location. The slight dip in $B^S_{DC}$ around 25 K is reproducible between cooldowns and suggests that the closest domains to our NV location are anti-aligned with the external field $H_{DC}$. The domains, however, can be polarized along the dc field at $H_{DC}$ = 76 mT (purple points). During subsequent FW, the magnetization, proportional to $-B^S_{DC}$, is well described by a critical model $M(T) \propto (1 - T/T_c)^\beta$ with $\beta = 0.38 \pm 0.02$ and $T_c = 30.5$ K. These results are in close agreement with prior works which found $\beta \approx 0.4$ for both bulk \cite{Ho1969} and monolayer \CrBr{} \cite{Kim2019a}.

In contrast to its vanishing dc magnetization, the ac susceptibility \Xac{} for Flake C reveals distinct features that highlight the formation and mobility of domains (Fig. \ref{fig:4}a, blue points). As \Tc{} is approached, the paramagnetic divergence of \Xac{} is stunted, and \Xac{} decreases initially, leading to a local maximum slightly below \Tc{} (Hopkinson peak).  Monte Carlo simulations \cite{Wahab2020,Tiwari2020} indicate that incipient domain-like patches begin to form a few Kelvin above \Tc{}, with the magnetic axes of the domains initially isotropically distributed. We attribute the initial rounding and decrease of \Xac{} to the formation of this domain structure, which impedes coherent MR relative to the paramagnetic state.

As temperature is reduced further, the measured \Xac{} for Flake C starts to steadily increase (Fig. \ref{fig:4}a). Here, the magnetic anisotropy is increasing, and we expect a crossover to a 180$^\circ$ domain phase with magnetizations fully parallel or anti-parallel to the easy ($z$-) axis \cite{Wahab2020,Tiwari2020}. Accordingly, \Xac{} becomes dominated by DWD, while the MR contribution is phased out. For DWDs described by the membrane-like bulging of a domain wall between two pinned endpoints, the susceptibility $\chi_{DW}$ is proportional to $M_S^2 l / \gamma_{DW}$, where $M_S$ is the saturation magnetization, $l$ is the length between the pinned ends, and $\gamma_{DW} \propto \sqrt{A_{ex} K}$ is the domain wall energy \cite{Chikazumi1997}. Here, $A_{ex}$ is the exchange stiffness parameter and $K$ is the uniaxial anisotropy. Making the approximations $A_{ex} \propto M_S^2$ and $K \propto M_S^3$ for bulk \CrBr{} \cite{Richter2018}, the factor $M_S^2 / \gamma_{DW}$ $(\propto M_S^{-1/2})$ would decrease with decreasing temperature. Thus, the increasing \Xac{} observed experimentally instead suggests that the length $l$ of the domain wall that can be continuously displaced is increasing for a range of decreasing temperatures, assuming constant domain wall density \cite{Bostanjoglo1970a}. This could reflect the change in the domain morphology observed in imaging experiments on bulk \CrBr, where smoother, straighter striped domains evolve at low temperature from zigzag stripes near \Tc{} with short interlocking sections \cite{Bostanjoglo1970a,Kuhlow1975}.
 
We now present \Xac{} measurements during the initial magnetization process for Flake A, which in contrast to Flake C, displays a rectangular hysteresis curve at 4 K (Fig. \ref{fig:2}d), characteristic of ultrathin (e.g. monolayer) flakes. This indicates that domain walls near the NV probe location can propagate freely once nucleated, allowing rapid formation of large domains. We hypothesize that Flake A's domain morphology resembles the disordered, patchy domains visualized by scanning NV magnetometry for three to four layer \CrBr{} \cite{Sun2020}, which also displayed rectangular hysteresis, rather than the periodic, narrow stripe patterns seen in bulk \CrBr{}, where hysteresis and dc stray field are absent (e.g. Flake C).

For FC at low $H_{DC}$ = 2 mT (Fig. \ref{fig:4}b), we see a much stronger divergence of the paramagnetic susceptibility in absolute units for Flake A (3 times stronger than Flake C), suggesting reduced interference from a periodic domain structure. This allows us to extract $T_c = 30.5 \pm 0.5$ K from the inflection point in \Xac{} versus temperature and a critical exponent $\gamma = 1.1 \pm 0.3$ in the scaling of $\chi_{AC}(T) \propto (T/T_c-1)^\gamma$ above \Tc{}. This value of $\gamma$ signals that Flake A is still proximal to the bulk, mean-field limit of magnetic interactions ($\gamma = 1$) \cite{Ho1969}. However, its detailed domain dynamics below \Tc{}, as revealed through ac susceptibility, differs strongly from the thicker Flake C, also pointing to a different underlying domain morphology.

For Flake A, we observe only a small kink in \Xac{} at 29 K, rather than the broad Hopkinson peak seen for Flake C. \Xac{} reaches a maximum at an intermediate temperature (${\sim}$14 K), before decreasing to a small value at 4 K. The high peak value of \Xac{} corroborates the enhanced domain wall mobility suggested by the rectangular hysteresis curve. The eventual decrease of \Xac{} at lower temperatures reflects the combination of two factors. First, the net area of domain walls near the NV location decreases as the magnetic domains anti-parallel to $H_{DC}$ shrink or disappear to minimize the domain wall energy \cite{Wahab2020}. This process is reflected by the increase in the dc magnetization of the flake during FC (Fig. \ref{fig:4}b, red data). Second, the mobility for the remaining domain walls is reduced due to increasing magnetic anisotropy $K$ at lower temperatures, which will be corroborated by frequency-dependent measurements.

To gain further insight, we examine how features in the temperature dependence of \Xac{} for Flake A evolve with increasing bias field $H_{DC}$ during FC. As shown in Fig. \ref{fig:4}b-d, the behavior of \Xac{} for different $H_{DC}$ closely overlaps in the paramagnetic region, but separates below \Tc{}. For larger $H_{DC}$, the enhanced field pressure $p = 2M_S H_{DC}$ accelerates the depinning and expansion of domains aligned with the field, causing a more rapid rise in the dc magnetization. Concomitantly, \Xac{} peaks at higher temperatures during FC and terminates at lower values at 4 K, reflecting accelerated domain consolidation and reduction of domain wall area.

These effects are particularly dramatic for FC at the highest $H_{DC}$ = 10.6 mT (Fig. \ref{fig:4}d), where \Xac{} is significantly reduced below $T_c$ as the flake immediately forms large domains. The dual-humped shape of \Xac{} should reflect the motion of only a handful of domain walls in the pinning landscape of Flake A, since the dc magnetization reaches approximately the saturated, single-domain value and \Xac{} drops to zero at 4 K. Away from \Tc, \Xac{} should evolve smoothly when averaging over the behavior of a statistical ensemble of domain walls. In contrast, the sharp turnaround of \Xac{} here at ${\sim}$20 K can be interpreted to correspond to the thermal depinning of a single domain wall from a strong pinning center, analogous to a Barkhausen jump in the dc hysteresis measurement.

\section{Frequency-Dependent AC Susceptibility}
Finally, we demonstrate frequency-dependent ac susceptibility measurements by sweeping $f_{AC}$ of the coil field, while changing the repetition rate of the NV $\pi$-pulses to maintain lock-in condition. We increase the number of $\pi$-pulses for larger $f_{AC}$ so that the total duration of the sensing sequence remains roughly constant, giving us similar precision on the determination of \Xac{}. For $H_{DC} =$ 6 mT, we gather data for $f_{AC}$ spanning 119 kHz (using 8 $\pi$-pulses) to 714 kHz (48 $\pi$-pulses). The lower range is limited by the $^{13}$C nuclear spin bath inside diamond, which reduces the NV center coherence as $f_{AC}$ overlaps the $^{13}$C Larmor frequency (64 kHz), and the upper by the finite duration of our $\pi$-pulse. Intrinsically, NV susceptometry is applicable to a wider frequency range, leveraging longer coherence times in surface-engineered, isotopically-purified samples \cite{Eichhorn2019} and faster spin inversions in the strong driving regime ($>$100 MHz effective Rabi frequency \cite{Fuchs2009}).

\begin{figure}[t]
\includegraphics[scale=1]{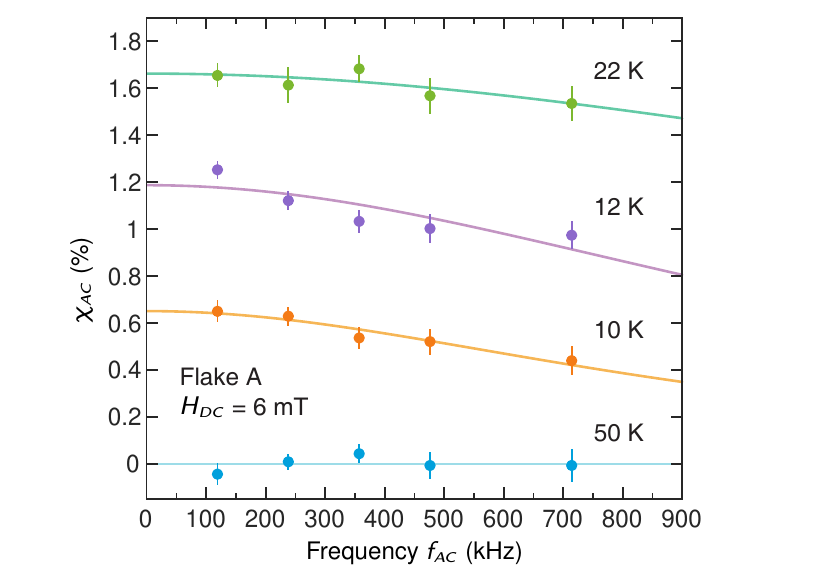}
\caption{\label{fig:5}Frequency-dependent susceptibility of domain walls in ultrathin \CrBr{}. \Xac{} for Flake A at 50 K (blue), 22 K (green), 12 K (purple), and 10 K (orange) during FC. The solid lines are fits using a damped, driven domain wall oscillation model. The number of $\pi$-pulses increases for larger $f_{AC}$ (e.g., XY8-$N6$ is used for 714 kHz).}
\end{figure}

As displayed in Fig. \ref{fig:5}, \Xac{} for Flake A is independent of $f_{AC}$ well above \Tc{} (50 K, blue), as expected. At 22 K, however, during the initial rise of the field-cooling \Xac{} (green arrow in Fig. \ref{fig:4}c), we observe a weak decrease in \Xac{} for higher excitation frequencies. This effect becomes more pronounced upon further cooling, as \Xac{} begins to decrease with temperature. Increasing $f_{AC}$ sixfold from 119 kHz to 714 kHz, the susceptibility \Xac{} is reduced by a factor of $0.90\pm0.08$ at 22 K, $0.78\pm0.05$ at 12 K, and $0.68\pm0.10$ at 10 K.

As first elucidated by D\"{o}ring \cite{Rado1950a}, the displacement $s$ of a domain wall satisfies the phenomenological equation:
\begin{equation}\label{eq:1}
m\,d^2s/dt^2 + \beta\,ds/dt + \alpha\,s = 2 M_S H,
\end{equation}
where $m$ is the apparent mass of the wall, $\beta$ is the viscous damping, $\alpha$ is the restoring force due to the local potential, and $2 M_S H$ is the driving force due to the field pressure on a 180$^\circ$ domain wall. For an oscillating field $H = H_0 e^{i \omega t}$ with $\omega = 2\pi f_{AC}$, the susceptibility $\chi_{DW}$ is proportional to the displacement $s$ and is thus analogous to the transfer function of a driven damped harmonic oscillator:
\begin{equation}\label{eq:2}
\chi_{DW} = \frac{\chi_0}{1-(\omega/\omega_0)^2 + i (\omega/\omega_c)}
\end{equation}
where $\chi_0$ is the dc susceptibility, $\omega_0 = \sqrt{\alpha/m}$ is the domain wall resonance, and $\omega_c = \alpha/\beta$ is a relaxation frequency, reflecting the timescale $\tau_c = \beta/\alpha$ for displacements to decay. For dc fields $H$, the phenomenological mobility $\mu = (ds/dt)/H$ of a uniform domain wall is $2M_S/\beta$ in this model \cite{Guyot1982}.

The domain wall resonance $\omega_0$ for thin \CrBr{} platelets has been measured to exceed 700 MHz \cite{Jedryka1982}, and hence $(\omega/\omega_0)^2$ can be neglected for excitation frequencies applied here, leaving
\begin{equation}\label{eq:3}
\operatorname{Re}(\chi_{DW}) = \frac{\chi_0}{1+(\omega/\omega_c)^2}.
\end{equation}
The solid lines in Fig. \ref{fig:5} show fits of our data to Eq. \ref{eq:3}. The fitted $\omega_c/2\pi$ decreases monotonically from ${\sim}$2 MHz at 22 K to ${\sim}$1 MHz at 10 K, indicating longer relaxation times resulting most likely from stronger domain wall damping $\beta$. For ferromagnetic insulators, $\beta$ is contributed by spin-lattice relaxation that similarly damps magnetization precession (Gilbert damping) and is proportional to $M_S\sqrt{K/A}$, which increases with decreasing temperature \cite{Infante2009}. In context, the robustness of \Xac{} here to frequencies exceeding hundreds of kilohertz contrasts with the behavior of conventional ferromagnetic materials that are commonly characterized in SQUID magnetometers with bandwidths of a few kilohertz. Our measurements thus underscore the overall low damping and high mobility of domain walls in ultrathin \CrBr{} \cite{Wahab2020}.

\section{Conclusion}
The ac susceptometry technique developed here establishes NV magnetometry as a true multi-modal probe for ultrathin magnetic materials, combining nanoscale spatial resolution and quantitative magnetization characterization previously demonstrated using dc field measurements \cite{Thiel2019,Broadway2020a,Sun2020,Fabre2020} with now access to a wide bandwidth of dynamical magnetization phenomena. Although we did not measure monolayer samples, the precision on our ac field measurements (${\sim}$40 nT), enabled by our active protection of the NV spin coherence, exceeds that of prior dc magnetometry on 2D magnetic monolayers by two orders of magnitude. This remarkable sensitivity allowed us to resolve the ac magnetic response due to paramagnetic spin rotation and domain wall motion in single, few-layer \CrBr{} flakes, showing the latter to be strongly influenced by the domain morphology and local pinning landscape of the flake. A future opportunity lies in the investigation of the out-of-phase component of the ac susceptibility, which can be compared through the fluctuation-dissipation theorem to simultaneous spin noise spectra measured by NV magnetometry \cite{Romach2015}. Our exciting development thus opens the door to understanding sub-gigahertz magnetic dynamics in diverse 2D materials, including antiferromagnets \cite{Wang2019b}, superconductors \cite{Xi2016}, and quantum spin liquids \cite{Xu2020} that lack magneto-optical coupling, and in single molecule magnets \cite{Guo2018} and superparamagnetic nanoparticles \cite{VanDeLoosdrecht2019} at the single particle limit.

\section{Acknowledgments}
The authors thank B. Flebus and Q. Ma for valuable discussions. B.B.Z acknowledges support from the National Science Foundation award No. ECCS-2041779. F.T. acknowledges support from the National Science Foundation award No. DMR-1708929. M.J.G. was supported by the National Science Foundation grant DMR-2003343, and K.S.B. was supported by the Office of Naval Research under award No. N00014-20-1-2308.

X.-Y.Z. and B.B.Z. devised the experiment. X.-Y.Z. and Y.-X.W. built the cryogenic confocal system. X.-.Y.Z. acquired and analyzed the data with assistance from Y.-X.W., T.D., and B.B.Z. T.A.T and F.T. synthesized the \CrBr{} samples. M.J.G. and K.S.B maintained the glovebox and AFM, and instructed X.-Y.Z. on their use. B.B.Z. and X.-Y.Z. wrote the paper with contributions from all authors.

\newpage 
%

\end{document}